\documentclass{article}
\usepackage{epsfig}
\usepackage{graphics}

\renewcommand{\arraystretch}{1.3}
\makeatletter
\newdimen\normalarrayskip              
\newdimen\minarrayskip                 
\normalarrayskip\baselineskip \minarrayskip\jot
\newif\ifold             \oldtrue            \def\new{\oldfalse}
\def\arraymode{\ifold\relax\else\displaystyle\fi} 
\def\eqnumphantom{\phantom{(\theequation)}}     
\def\@arrayskip{\ifold\baselineskip\z@\lineskip\z@
     \else
     \baselineskip\minarrayskip\lineskip2\minarrayskip\fi}
\def\@arrayclassz{\ifcase \@lastchclass \@acolampacol \or
\@ampacol \or \or \or \@addamp \or
   \@acolampacol \or \@firstampfalse \@acol \fi
\edef\@preamble{\@preamble
  \ifcase \@chnum
     \hfil$\relax\arraymode\@sharp$\hfil
     \or $\relax\arraymode\@sharp$\hfil
     \or \hfil$\relax\arraymode\@sharp$\fi}}
\def\@array[#1]#2{\setbox\@arstrutbox=\hbox{\vrule
     height\arraystretch \ht\strutbox
     depth\arraystretch \dp\strutbox
     width\z@}\@mkpream{#2}\edef\@preamble{\halign
\noexpand\@halignto
\bgroup \tabskip\z@ \@arstrut \@preamble \tabskip\z@ \cr}%
\let\@startpbox\@@startpbox \let\@endpbox\@@endpbox
  \if #1t\vtop \else \if#1b\vbox \else \vcenter \fi\fi
  \bgroup \let\par\relax
  \let\@sharp##\let\protect\relax
  \@arrayskip\@preamble}
\def\eqnarray{\stepcounter{equation}%
              \let\@currentlabel=\theequation
              \global\@eqnswtrue
              \global\@eqcnt\z@
              \tabskip\@centering
              \let\\=\@eqncr
 \halign to \displaywidth\bgroup
    \eqnumphantom\@eqnsel\hskip\@centering
    $\displaystyle \tabskip\z@ {##}$%
    \global\@eqcnt\@ne \hskip 2\arraycolsep
         $\displaystyle\arraymode{##}$\hfil
    \global\@eqcnt\tw@ \hskip 2\arraycolsep
         $\displaystyle\tabskip\z@{##}$\hfil
         \tabskip\@centering
    &{##}\tabskip\z@\cr}
\begingroup\ifx\undefined\newsymbol \else\def\input#1 {\endgroup}\fi

\catcode`\@=11
\def\marginnote#1{}

\newcount\hour
\newcount\minute
\newtoks\amorpm
\hour=\time\divide\hour by60 \minute=\time{\multiply\hour by60
\global\advance\minute by-\hour}
\edef\standardtime{{\ifnum\hour<12 \global\amorpm={am}%
        \else\global\amorpm={pm}\advance\hour by-12 \fi
        \ifnum\hour=0 \hour=12 \fi
        \number\hour:\ifnum\minute<10 0\fi\number\minute\the\amorpm}}
\edef\militarytime{\number\hour:\ifnum\minute<10 0\fi\number\minute}

\def\draftlabel#1{{\@bsphack\if@filesw {\let\thepage\relax
      \xdef\@gtempa{\write\@auxout{\string
          \newlabel{#1}{{\@currentlabel}{\thepage}}}}}\@gtempa \if@nobreak
    \ifvmode\nobreak\fi\fi\fi\@esphack} \gdef\@eqnlabel{#1}}
    \def\@eqnlabel{}
\def\@vacuum{}
\def\draftmarginnote#1{\marginpar{\raggedright\scriptsize\tt#1}}

\def\draft{
  \oddsidemargin -.5truein
  \def\@oddfoot{\footnotesize \sl preliminary draft \hfil
    \rm\thepage\hfil\sl\today\quad\militarytime}
  \let\@evenfoot\@oddfoot \overfullrule 3pt
    \let\label=\draftlabel
    \let\marginnote=\draftmarginnote
  \def\@eqnnum{(\theequation)\rlap{\kern\marginparsep\tt\@eqnlabel}%
    \global\let\@eqnlabel\@vacuum}

  }

\def\be{\begin{eqnarray}}
\def\ee{\end{eqnarray}}
\def\nn{\nonumber}

\def\beq{\begin{equation}}
\def\eeq{\end{equation}}
\def\ba{\beq\new\begin{array}{c}}
\def\ea{\end{array}\eeq}
\def\be{\ba}
\def\ee{\ea}

\newfont{\alef}{msbm10 at 12pt}
\newfont {\goth}{eufm10 at 11pt}
\def\mathbb#1{\hbox{{\alef #1}}}

\let\@@savethanks\thanks
\def\thanks#1{\gdef\thefootnote{\alph{footnote}}\@@savethanks{#1}}

\title{{\bf Some properties of the Alday-Maldacena minimum
} \vspace{.5cm}}
\author{{\bf A. Mironov}\footnote{E-mail: \ mironov@itep.ru; mironov@lpi.ru}
\date{ } \\
{\small {\it Lebedev Physics Institute}
and {\it ITEP, Moscow, Russia}}\\ \\
{\bf A. Morozov}\thanks{E-mail: \ morozov@itep.ru}
\date{ } \\ {\small {\it ITEP, Moscow, Russia}}
\\ \\
{\bf T.N. Tomaras}\thanks{E-mail: \ tomaras@physics.uoc.gr}
\date{ } \\ {\small {\it Department of Physics and Institute of Plasma Pysics, University of Crete, Heraklion; Greece}}
}

\begin{document}

\maketitle

\vspace{-10.5cm}

\begin{center}
\hfill FIAN/TD-19/07\\
\hfill ITEP/TH-39/07\\
\end{center}

\vspace{9.0cm}

\begin{abstract}
\noindent The Alday-Maldacena solution, relevant to the $n=4$ gluon
amplitude in $N=4$ SYM at strong coupling, was recently identified
as a minimum of the regularized action in the moduli space of
solutions of the $AdS_5$ $\sigma$-model equations of motion.
Analogous solutions of the Nambu-Goto equations for the $n=4$ case
are presented and shown to form (modulo the reparametrization group)
an equally large but different moduli space, with the
Alday-Maldacena solution at the intersection of the $\sigma$-model
and Nambu-Goto moduli spaces. We comment upon the possible form of
the regularized action for $n=5$. A function of moduli parameters
$z_a$ is written, whose minimum reproduces the BDDK one-loop
five-gluon amplitude. This function may thus be considered as some
kind of Legendre transform of the BDDK formula and has its own value
independently of the Alday-Maldacena approach.
\end{abstract}

\bigskip

\def\thefootnote{\arabic{footnote}}

\bigskip

\bigskip

\section{Introduction and conclusions}

An $\epsilon$-regularized minimal action in
$AdS_5$ $\sigma$-model was defined recently \cite{AM}, and shown to reproduce
the external momentum dependence of the BDS formula \cite{BDS} for
the $n=4$-gluon amplitude in $N=4$ Super Yang-Mills (SYM) theory. In
\cite{AMoth}-\cite{Popo} one may find generalizations and
discussions of this important result. In a previous paper
\cite{MMT} it was demonstrated that the Alday-Maldacena solution is just
one member of a large family of solutions; a rather distinct one, though, since it corresponds
to a minimum of the classical $\sigma$-model action in the moduli
space ${\cal M}_n^{\sigma}$ of all solutions in $d=4$ dimensions (i.e. for
$\epsilon=0$). Throughout this paper
we shall use the notation and results of \cite{MMT}, to which we refer the reader.
We shall keep the parameter $n$ explicit in various formulas and symbols,
even though, as it will be clear in the text, many of the statements will refer specifically
to the cases $n=4$ or $5$.

\begin{figure}
\hspace{2.5cm}
\centerline{
\unitlength 1mm 
\linethickness{0.4pt}
\ifx\plotpoint\undefined\newsavebox{\plotpoint}\fi 
\begin{picture}(160.874,79.78)(0,0)
\put(66.705,20.348){\circle{.867}}
\put(67.392,76.808){\circle{.867}}
\put(19.017,58.808){\circle{.867}}
\put(18.497,55.625){\circle{.867}}
\put(17.828,52.416){\circle{.867}}
\put(71.238,21.537){\circle{.867}}
\put(72.787,76.072){\circle{.867}}
\put(75.593,22.429){\circle{.867}}
\put(77.557,75.231){\circle{.867}}
\multiput(87.157,29.974)(-.03363,-.03994){100}{\line(0,-1){.03994}}
\multiput(83.794,25.98)(.0315,-.0946){10}{\line(0,-1){.0946}}
\multiput(84.109,25.034)(.10957447,.03355319){47}{\line(1,0){.10957447}}
\multiput(89.259,26.611)(-.1616923,.0323077){13}{\line(-1,0){.1616923}}
\multiput(87.157,27.031)(.0500823529,.0336970588){340}{\line(1,0){.0500823529}}
\multiput(104.185,38.488)(-.1131538,.0323846){13}{\line(-1,0){.1131538}}
\multiput(102.714,38.909)(-.03355319,.09391489){47}{\line(0,1){.09391489}}
\multiput(101.137,43.323)(.0663684,-.0331579){19}{\line(1,0){.0663684}}
\multiput(102.398,42.693)(-.03343939,.30577273){66}{\line(0,1){.30577273}}
\multiput(100.191,62.874)(-.08408333,.03328333){60}{\line(-1,0){.08408333}}
\multiput(95.146,64.871)(.1997,.0316){10}{\line(1,0){.1997}}
\multiput(97.143,65.187)(-.047186667,.033635556){225}{\line(-1,0){.047186667}}
\multiput(86.526,72.755)(.0334545,.0382273){22}{\line(0,1){.0382273}}
\multiput(35.777,76.005)(.0334545,.0382273){22}{\line(0,1){.0382273}}
\multiput(22.152,28.63)(.0334545,.0382273){22}{\line(0,1){.0382273}}
\multiput(35.902,18.13)(.0334545,.0382273){22}{\line(0,1){.0382273}}
\multiput(87.262,73.596)(.045927928,-.033617117){222}{\line(1,0){.045927928}}
\put(97.458,66.133){\line(0,1){1.366}}
\multiput(97.353,67.499)(.033525862,-.035336207){116}{\line(0,-1){.035336207}}
\multiput(101.242,63.4)(.03336508,-.32868254){63}{\line(0,-1){.32868254}}
\multiput(103.344,42.693)(.03364,.06304){25}{\line(0,1){.06304}}
\put(104.185,44.269){\line(0,-1){5.781}}
\multiput(100.086,63.084)(.0889231,.0323846){13}{\line(1,0){.0889231}}
\put(86.83,73.23){\circle*{1.863}}
\put(36.081,76.48){\circle*{1.863}}
\put(22.456,29.105){\circle*{1.863}}
\put(36.206,18.605){\circle*{1.863}}
\put(100.729,63.196){\circle*{1.863}}
\put(103.479,38.743){\circle*{1.863}}
\put(83.708,25.067){\circle*{1.863}}
\multiput(20.75,62.875)(.0402822096,.0336377214){301}{\line(1,0){.0402822096}}
\multiput(32.875,73)(-.03125,-.34375){4}{\line(0,-1){.34375}}
\multiput(32.75,71.625)(.033601998,.044354638){93}{\line(0,1){.044354638}}
\multiput(53.875,75.875)(-.0326085,-.0326085){23}{\line(0,-1){.0326085}}
\multiput(53.125,75.125)(.16158463,.03353643){41}{\line(1,0){.16158463}}
\multiput(59.75,76.5)(-.12499943,.03365369){52}{\line(-1,0){.12499943}}
\multiput(53.25,78.25)(.03125,-.072916){12}{\line(0,-1){.072916}}
\multiput(53.625,77.375)(-2.203115,-.03125){8}{\line(-1,0){2.203115}}
\multiput(36,77.125)(-.059139517,-.033601998){93}{\line(-1,0){.059139517}}
\multiput(30.5,74)(.375,-.03125){4}{\line(1,0){.375}}
\multiput(32,73.875)(-.0409554459,-.0337029191){293}{\line(-1,0){.0409554459}}
\multiput(20,64)(.0326085,-.0489128){23}{\line(0,-1){.0489128}}
\multiput(40.625,17.5)(.0333332,-.0374998){30}{\line(0,-1){.0374998}}
\multiput(41.625,16.375)(-.14144673,.03289459){38}{\line(-1,0){.14144673}}
\multiput(36.25,17.625)(-.0443875541,.0336733169){245}{\line(-1,0){.0443875541}}
\put(25.375,25.875){\line(0,-1){1.625}}
\multiput(25.375,24.25)(-.03341569,.044554254){101}{\line(0,1){.044554254}}
\multiput(22,28.75)(-.033601998,.139784313){93}{\line(0,1){.139784313}}
\multiput(18.875,41.75)(-.0326085,-.0489128){23}{\line(0,-1){.0489128}}
\put(18.125,40.625){\line(0,1){6.5}}
\multiput(18.125,47.125)(.03343008,-.06104624){86}{\line(0,-1){.06104624}}
\multiput(21,41.875)(-.114583,.03125){12}{\line(-1,0){.114583}}
\put(160.874,29.25){\line(0,1){0}}
\multiput(19.625,42.125)(.033601998,-.137096153){93}{\line(0,-1){.137096153}}
\multiput(25.75,26.75)(.045405777,-.033653694){234}{\line(1,0){.045405777}}
\put(36.375,18.875){\line(5,1){5.625}}
\multiput(42,20)(-.03353643,-.03353643){41}{\line(-1,0){.03353643}}
\put(40.625,18.625){\line(1,0){16.125}}
\put(56.75,18.625){\line(0,-1){1.125}}
\multiput(25.625,26.75)(.145833,.03125){12}{\line(1,0){.145833}}
\multiput(27.375,27.125)(-.06902954,.03358194){67}{\line(-1,0){.06902954}}
\multiput(36.125,75.875)(2.21874,.03125){8}{\line(1,0){2.21874}}
\multiput(87.125,29.875)(-.03125,-.145833){12}{\line(0,-1){.145833}}
\multiput(86.75,28.125)(.0503965972,.033730006){315}{\line(1,0){.0503965972}}
\multiput(40.625,17.625)(4.03123,.03125){4}{\line(1,0){4.03123}}
\put(36.25,18.875){\line(0,1){57.5}}
\multiput(36.25,76.375)(-.033665683,-.1181416097){401}{\line(0,-1){.1181416097}}
\thicklines
\multiput(34.125,70.625)(.03348199,.09151744){56}{\line(0,1){.09151744}}
\multiput(36,75.75)(-.0336537,-.2067298){26}{\line(0,-1){.2067298}}
\put(87.157,76.118){\makebox(0,0)[cc]{$n$}}
\put(95.041,71.388){\makebox(0,0)[cc]{${\bf p}_n$}}
\put(103.344,64.556){\makebox(0,0)[cc]{$1$}}
\put(105.131,52.994){\makebox(0,0)[cc]{${\bf p}_1$}}
\put(106.287,37.647){\makebox(0,0)[cc]{$2$}}
\put(96.933,29.659){\makebox(0,0)[cc]{${\bf p}_2$}}
\put(84.845,22.406){\makebox(0,0)[cc]{$3$}}
\put(6.185,60.158){\makebox(0,0)[cc]{${\bf P}_{ab}$}}
\put(70.339,50.891){\makebox(0,0)[cc]{$t_{ab}=({\bf p}_a+{\bf P}_{ab})^2$}}
\put(45.408,79.78){\makebox(0,0)[cc]{${\bf p}_b$}}
\put(33.951,79.359){\makebox(0,0)[cc]{$b$}}
\put(14.295,36.263){\makebox(0,0)[cc]{${\bf p}_{a+1}$}}
\put(17.133,28.38){\makebox(0,0)[cc]{$a+1$}}
\put(27.434,20.917){\makebox(0,0)[cc]{${\bf p}_a$}}
\put(34.161,15.557){\makebox(0,0)[cc]{$a$}}
\put(47.826,14.821){\makebox(0,0)[cc]{${\bf p}_{a-1}$}}
\thinlines
\qbezier(11.142,60.439)(13.77,60.649)(17.238,61.28)
\put(30.798,59.283){\vector(4,-1){.07}}\qbezier(17.238,61.28)(20.97,61.648)(30.798,59.283)
\qbezier(54.238,51.4)(49.455,51.452)(46.144,47.931)
\put(36.999,45.408){\vector(-1,0){.07}}\qbezier(46.144,47.931)(43.148,45.618)(36.999,45.408)
\thicklines
\multiput(34.897,70.74)(.0328474,.1445284){32}{\line(0,1){.1445284}}
\multiput(35.948,75.365)(-.03332806,-.09998418){41}{\line(0,-1){.09998418}}
\multiput(34.582,71.055)(.0328474,.111681){32}{\line(0,1){.111681}}
\multiput(35.633,74.629)(-.03319313,-.09681329){38}{\line(0,-1){.09681329}}
\end{picture}}
\caption{{\footnotesize
Auxiliary skew polygon $\Pi$, playing a surprisingly important
role in the theory of $n$-point amplitudes: all formulas at the
perturbative, as well as the strong-coupling sides of the AdS/CFT correspondence
are written in terms of characteristics of $\Pi$.
Its edges are external gluon 4-momenta ${\bf p}_a$, the squares of its
diagonals are scattering invariants $t_{ab}$.
Formulas in the text are written in terms of
their logarithms, $\tau_{ab} = \log t_{ab}$.
}}
\label{polyg}
\end{figure}
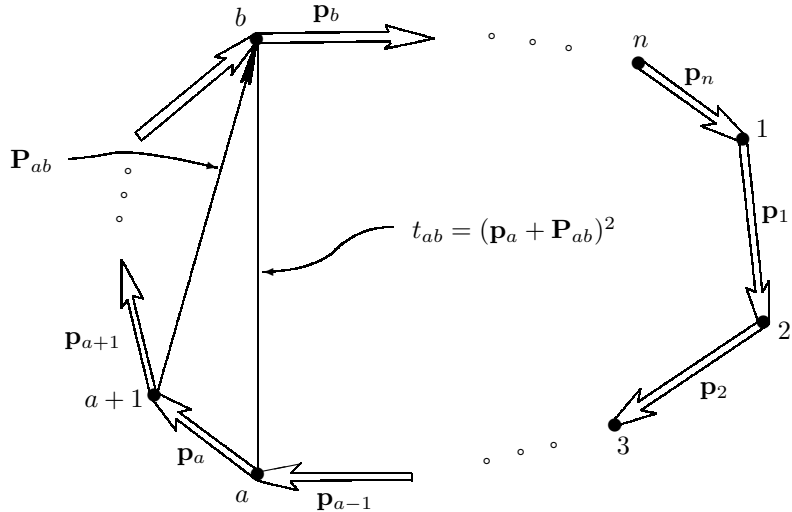

Let us recall that in $d=4$ dimensions and for $n=4$ the moduli space of solutions
constructed in \cite{MMT} was parametrized by $\{z_a,{\bf v_1},\phi\}$ with
$a=1, \ldots , n$ enumerating the sides of the auxiliary polygon $\Pi$,
Figure \ref{polyg}, formed by the null 4-momenta ${\bf p}_a$ of the external
gluons and lying at the boundary of $AdS_5$ at $z=\infty$.

It is possible, that these are all the solutions with the particular boundary
conditions corresponding to the above process. In \cite{MMT} they were obtained
under the assumption (ansatz) that the lagrangian $L_\sigma = {\rm constant} = 2$.
In any case, in what follows we shall use ${\cal M}_n^\sigma$ to denote {\it this} part of the
moduli space.

The $SO(4,2)$ symmetry of $AdS_5$ relates some of these solutions,
but it does not act transitively on ${\cal M}_n^\sigma$. Specifically, only
the $z_a$ moduli are affected by this group. In addition, ${\bf v}_1$ is an
inessential modulus, since no physical quantity depends on it. Essential
moduli are the ratio $z_1z_3/z_2z_4$ and the angle $\phi$. The latter is
not affected by $SO(4,2)$, but only by some larger hidden group,
related presumably to the integrability of the $\sigma$-model. It is important to point out that
by definition the Lagrangian density is constant, namely $L_\sigma=2$, on the entire ${\cal M}_n^\sigma$.
Thus, the corresponding action integral diverges and needs regularization. The $\epsilon$-regularization used in \cite{AM}
breaks not only the integrability, but also the $SO(4,2)$ symmetry,
so that the regularized action becomes
a non-trivial ($z$ and $\phi$-dependent) function on the moduli
space. As shown in \cite{MMT}, the Alday-Maldacena solution is
exactly at the minimum of this function. Incidentally, the regularization leaves
unbroken the Lorentz subgroup of $SO(4,2)$ (which, however, is partly broken by the
boundary conditions) and the two rescalings of $z_a$ that preserve the
products $z_1z_3$ and $z_2z_4$.

The present note is a little further development along the lines of \cite{MMT}. Our purpose is
on the one hand to clarify the difference between the $\sigma$-model and Nambu-Goto actions
in connection with the above approach, and on the other to attempt a generalization
to the five-gluon amplitude.

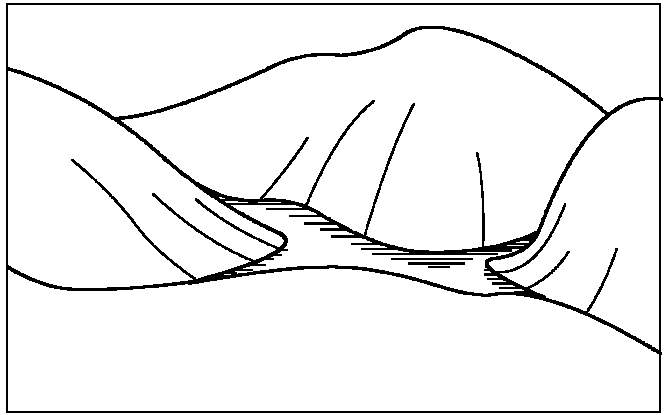
\begin{figure}
\centerline{
\unitlength 1mm 
\linethickness{0.4pt}
\ifx\plotpoint\undefined\newsavebox{\plotpoint}\fi 
\begin{picture}(97.545,61.678)(0,0)
\put(10.568,61.678){\line(1,0){86.85}}
\put(97.418,61.678){\line(0,-1){54.3}}
\put(97.418,7.378){\line(-1,0){86.85}}
\put(10.568,7.378){\line(0,1){54.15}}
\thicklines
\qbezier(10.783,53.033)(22.398,49.533)(31.678,40.941)
\qbezier(24.572,46.457)(30.459,46.298)(44.618,52.715)
\qbezier(44.618,52.715)(49.657,55.473)(54.907,54.836)
\qbezier(54.907,54.836)(59.839,55.048)(63.71,57.806)
\qbezier(63.71,57.806)(67.9,60.192)(77.817,55.154)
\qbezier(77.817,55.154)(84.605,51.866)(90.545,46.881)
\qbezier(97.545,49.002)(88.265,50.328)(81.317,31.502)
\qbezier(81.317,31.502)(79.885,28.267)(75.272,27.789)
\qbezier(75.272,27.789)(71.665,26.994)(82.06,22.592)
\qbezier(97.439,15.167)(82.325,24.448)(74.635,22.91)
\qbezier(74.635,22.91)(71.718,22.91)(67.741,24.183)
\qbezier(67.741,24.183)(58.725,26.994)(52.255,26.623)
\qbezier(52.255,26.623)(47.164,26.623)(34.86,24.501)
\qbezier(31.718,40.903)(39.331,34.453)(46.493,31.303)
\qbezier(46.493,31.303)(51.518,28.978)(34.643,24.553)
\qbezier(80.993,31.228)(74.693,28.303)(65.693,28.678)
\qbezier(65.693,28.678)(59.956,28.94)(51.968,33.853)
\qbezier(51.968,33.853)(48.106,36.065)(43.493,35.428)
\qbezier(43.493,35.428)(39.481,35.503)(35.918,37.678)
\thinlines
\qbezier(35.709,35.638)(39.952,31.926)(46.74,29.062)
\qbezier(43.77,27.259)(36.876,29.911)(29.981,36.381)
\qbezier(19.163,40.941)(24.837,36.275)(28.178,31.608)
\qbezier(28.178,31.608)(30.618,28.532)(35.603,25.032)
\thicklines
\qbezier(34.901,24.636)(25.759,23.521)(20.006,23.655)
\qbezier(20.006,23.655)(15.413,23.744)(10.641,26.687)
\thinlines
\qbezier(73.074,41.761)(74.1,37.346)(73.877,29.185)
\qbezier(64.691,48.361)(61.569,42.563)(58.09,30.701)
\qbezier(59.339,48.717)(54.567,44.971)(50.331,34.804)
\qbezier(50.509,43.812)(48.77,40.913)(44.355,35.696)
\qbezier(84.758,34.982)(81.191,25.572)(75.483,25.974)
\qbezier(85.204,28.65)(83.198,25.394)(79.585,23.744)
\qbezier(91.626,29.006)(90.065,24.324)(87.613,20.533)
\put(36.77,36.964){\line(1,0){.3712}}
\put(37.353,36.54){\line(1,0){.8485}}
\put(38.042,36.115){\line(1,0){1.3258}}
\put(38.944,35.532){\line(1,0){2.705}}
\put(39.792,35.055){\line(1,0){9.387}}
\put(57.771,29.062){\line(1,0){4.604}}
\put(59.222,28.369){\line(1,0){16.524}}
\put(61.682,27.675){\line(1,0){10.721}}
\put(63.889,27.107){\line(1,0){7.442}}
\put(66.664,26.666){\line(1,0){2.712}}
\put(74.169,26.666){\line(1,0){.378}}
\put(74.106,26.287){\line(1,0){.694}}
\put(74.043,25.657){\line(1,0){1.766}}
\put(74.484,25.089){\line(1,0){2.144}}
\put(75.178,24.459){\line(1,0){2.586}}
\put(76.061,23.828){\line(1,0){3.153}}
\put(78.079,23.26){\line(1,0){2.397}}
\put(80.412,22.819){\line(1,0){1.072}}
\put(79.593,30.576){\line(1,0){1.135}}
\put(78.268,30.198){\line(1,0){2.207}}
\put(76.881,29.756){\line(1,0){2.964}}
\put(75.43,29.252){\line(1,0){3.91}}
\put(72.781,28.873){\line(1,0){4.7931}}
\put(45.158,34.36){\line(1,0){5.928}}
\put(48.122,33.477){\line(1,0){4.289}}
\put(50.14,32.468){\line(1,0){4.099}}
\put(52.221,31.648){\line(1,0){3.532}}
\put(53.798,30.702){\line(1,0){4.0994}}
\put(56.321,29.756){\line(1,0){3.658}}
\put(47.807,28.936){\line(-1,0){1.324}}
\put(47.113,28.243){\line(-1,0){1.451}}
\put(45.978,27.675){\line(-1,0){1.514}}
\put(44.906,26.918){\line(-1,0){2.6488}}
\put(43.14,26.224){\line(-1,0){2.46}}
\put(40.995,25.846){\line(-1,0){1.324}}
\put(39.986,25.468){\line(-1,0){1.577}}
\put(38.473,25.152){\line(-1,0){1.4505}}
\end{picture}} \caption{\footnotesize Symbolic representation
of the landscape of world-sheet
embeddings into $AdS_5$ space. The horizontal plane is actually an
infinite-dimensional space of mappings $(z(\vec u),{\bf v}(\vec
u))$. The "height functionals" on this space are the
$\epsilon$-regularized actions. Actually the NG and $\sigma$-model
"height functions" are different, a fact ignored in this picture.
Solutions of the NG and $\sigma$-model equations of motion
form two valleys in this landscape. The $\sigma$-model one is not
flat, because the degeneracy is partly broken by the
$\epsilon$-regularization. Therefore, there is a minimum in the
valley, which coincides with the NG $\epsilon$-regularized action.
The Alday-Maldacena solution lies at the crossing of the two
valleys.} \label{cross}
\end{figure}

Specifically, in Section 2, we consider what happens if the $\sigma$-model action is
replaced by the Nambu-Goto (NG) one -- a question raised but
left unanswered in Section 4.7 of \cite{MMT}. We conclude that for $n=4$
the two moduli spaces are equally large, i.e.
\be
{\rm dim}\Big({\cal M}^\sigma_4\Big) =
{\rm dim}\Big({\cal M}_4^{NG}\Big)
\ee not expected a priori, because of the classical
inequivalence of the two actions. We would like to recall here, that the
$\sigma$-model is being considered without the Virasoro constraints, which would render
the two models classically equivalent. As shown, the relevant solutions are parametrized by the
same parameters, but they
are different in the two models and the corresponding moduli
spaces do not coincide. The essential moduli in the $\sigma$-model
case are the ratio $z_1z_3/z_2z_4$ and $\phi$, while in the NG case no
essential moduli are made from $z_a$. Instead, the angle $\phi$
between the two $\vec k$-vectors gets complemented by the ratio of
their lengths. This simple description, however, requires careful definition
of the manifold ${\cal M}_4^{NG}$. The reason is that, in contrast to the
$\sigma$-model case, the NG action is invariant under arbitrary
reparametrizations of the world sheet and, therefore, the entire space
of solutions is infinite dimensional, incomparably larger than that
of the $\sigma$-model.
In such a situation, it is natural to define the moduli
space by factoring out the reparametrization group with
coordinate transformations vanishing at infinity. Then the moduli
space of solutions with a given asymptotic behaviour at infinity is
finite-dimensional and is actually obtained by {\it linear}
transformations of the world-sheet coordinates. Similarly, it is natural to eliminate
the $2d$ rotations and displacements, since the $2d$ Poincare invariance is
common to the $\sigma$-model and NG actions. Next, the
$\epsilon$-regularization preserves the $2d$ reparametrization
invariance of the NG action, therefore, again in contrast to the
$\sigma$-model case, the regularized NG action is {\it constant} on
the entire ${\cal M}_4^{NG}$ manifold. The NG valley in the landscape of
world-sheet embeddings into $AdS_5$
is actually flat. It crosses the non-flat $\sigma$-model valley
exactly at the Alday-Maldacena solution
\footnote{For a small but potentially
interesting deviation see \cite{Popo}.}, Figure \ref{cross}.

In Section 3, guided by the pictorial representation given in \cite{MMT} and the results for $n=4$,
we make an attempt to guess the form of the $n>4$
regularized action ${\cal A}_n(z_1,\ldots,z_n;\epsilon)$ on the moduli space
${\cal M}_n^\sigma$, parametrized by a conjectured set of parameters $z_a$
with $a=1,2,...,n$. In addition, we present an ansatz for the constraint, generalization of
its $n=4$ counterpart, which is argued to be reasonable for $n=5$.
The action is minimized under the constraint and reproduces the BDDK formula\cite{BDDK} for the
one-loop 5-gluon amplitude $F_5^{(1)}$, which eventually exponentiates to the BDS formula \cite{BDS}
for the full strong coupling $n=5$ scattering amplitude. Hopefully, this action will eventually
be derived, as in the $n=4$ case, from exact solutions of the $\sigma$-model with subtle growing
asymptotics. At this point however, it may just serve as a useful guide through the tedious and not
particularly transparent calculations of the regularized integrals in \cite{AM} and \cite{MMT}.

It is important to emphasize here, that the finite part $\tilde{\cal A}_n$ of the $n$-point action
integral will be defined
independently of the Alday-Maldacena regularization. Consequently, it may be thought of
as a kind of Legendre transform of the BDDK formula and
in this sense has its own value and significance. Such a function for $n>5$
would have an advantage, because the $\{z_a\}$ are {\it independent} variables,
while there are many relations between the $n(n-3)/2$ parameters
$t_{ab}$, of which only $3n-10$ are independent. Construction of
$\tilde {\cal A}_n$ for $n>5$ is a challenging problem beyond the scope of the present paper.


\section{Moduli space of NG solutions for $n=4$}

In this section and in order to make the comparison easier,
we shall make a parallel presentation of the solutions of interest
in the Nambu-Goto (NG) and the $\sigma$-model field equations.

\subsection{Solving the NG equations of motion for $n=4$}

As explained in \cite{MMT}
the most relevant variables for the description of the Alday-Maldacena
result are $(z,{\bf v})$, which are actually five of the six flat
coordinates $(Y_-,{\bf Y},Y_+)$, describing the embedding of $AdS_5$
into $\mathbb{R}^6_{++++--}$. In these variables the equations of
motion acquire the simple form \be
\partial_i \Big(H^{ij}\partial_j z\Big)= G_{ij}H^{ij} z,
\label{eqm1}\ee\be
\partial_i \Big(H^{ij}{\bf V}_j\Big) = 0
\label{eqm2} \ee and the difference between the $\sigma$-model and
NG cases lies in the expression for $H$, namely, we have
\be H^{ij}_\sigma = \delta^{ij} \ee while \be
H^{ij}_{NG} = L_{NG} \Big(G^{-1}\Big)^{ij} = \frac{\check
G^{ij}}{L_{NG}} \ee In the above formulas $i,j=1,2$ label $2d$ coordinates on the
world sheet, \be G_{ij} = \frac{\partial_i z\partial_j z + {\bf
V}_i{\bf V}_j}{z^2} \label{Gmet} \ee is the $AdS$-induced metric on
the world sheet, $ \check G^{ij} = \left(\begin{array}{cc} G_{22} &
-G_{12}\\-G_{12}&G_{11}
\end{array}\right)
$ is made from algebraic complements, \be {\bf V} = z\partial{\bf v}
- {\bf v}\partial z \ee and the two lagrangian densities are \be
L_\sigma = G_i^i = G_{11}+G_{22} \ee and \be L_{NG} = \sqrt{\det G_{ij}}
= \sqrt{G_{11}G_{22}-G_{12}^{\,2}} \ee respectively.

In \cite{MMT} it was suggested to make the anzatz $G_{ij} = {\rm constant}
$ in the differential equations (\ref{eqm1}) and (\ref{eqm2}), solve
them with appropriate boundary conditions and finally consider the
self-consistency of this anzatz as an {\it algebraic} equation
(\ref{Gmet}) on the parameters of the solution. Many more NG solutions
can be produced afterwards by world sheet reparemeterizations $u^i
\rightarrow \tilde u^i(\vec u)$, corresponding to a single
point in the moduli space if $\ \tilde u^i = u^i +
O\big(|u|^{-1}\big)$ at large $|{\vec u}|$. For constant $G_{ij}$ both the
Lagrangian densities and the coefficient in front of $z$ on the
right hand side of  (\ref{eqm1}) are also constant, in which case the solutions of
equations (\ref{eqm1}), (\ref{eqm2}) are obviously of the form \be
z = \sum_a z_a e^{\vec k_a\vec u}, \nn\\
{\bf v} = \sum_a {\bf v}_ae^{\vec k_a\vec u} \label{exposol} \ee
where $\vec u$ are the world sheet coordinates. The $2d$ vectors
$\vec k_a$ are constrained in different ways for different actions:
\be \vec k_a^2 = {\rm Tr} G \ \ \ \ {\rm in \ the\ } \sigma-{\rm
model\ case}, \label{keqS} \ee\be \check G^{ij}k_i^ak_j^a =2 \det G
\ \ \ \ {\rm in\ the\ NG\ case} \label{keqNG} \ee Correspondingly, the parameters
${\bf v}_a$ are fixed by the boundary conditions \cite{MMT}, \be
\frac{{\bf v}_{a+1}}{z_{a+1}} - \frac{{\bf v}_a}{z_a} = {\bf p}_a
\ee which express them in terms of the external momenta ${\bf p}_a$ and
$\{z_a\}$. These boundary conditions restrict the number of
exponentials in (\ref{exposol}) to the number $n$ of sides in the
polygon $\Pi$: $a=1,\ldots,n$. One of the ${\bf v}$-vectors, say
${\bf v}_1$, remains undefined; we called it inessential modulus in
the introduction. The essential moduli are $\{z_a\}$ modulo $\vec u$
transformations and $\{\vec k_a\}$ {\rm modulo} (\ref{keqS}) {\rm
or} (\ref{keqNG}).

One is left with a set of non-trivial algebraic equations, namely that
$G_{ij}$ obtained by substitution of (\ref{exposol}) into
(\ref{Gmet}) is  constant, i.e. independent of $\vec u$: \be
\sum_{a,b=1}^n\Big(G_{ij}-k^a_ik^b_j\Big) z_az_b E_{a+b} =
\sum_{{a<b}\atop{c<d}}k^{ab}_ik^{cd}_j ({\bf{\cal P}}_{ab}{\bf{\cal
P}}_{cd}) z_az_bz_cz_dE_{a+b+c+d} \label{maineq} \ee Here $E_a =
e^{\vec k_a\vec u}$, $E_{a+b} = E_aE_b$, $\ \vec k^{ab}=\vec k^a -
\vec k^b$, $\ {\bf{\cal P}}_{ab} = z_az_b({\bf p}_a+\ldots+{\bf
p}_{b-1})$, while further details about notation can be found in
\cite{MMT}. In what follows we concentrate on the case of $n=4$,
where this simple ansatz indeed works. Equation (\ref{maineq}) is
actually a system of relations for coefficients in front of various
exponentials. Many coefficients can be cancelled if we choose $\vec
k_3=-\vec k_1 = \vec k_{-1}$ and $\vec k_4=-\vec k_2= \vec k_{-2}$
so that the four $\vec k$-vectors form diagonals of a parallelogram.
Next, comparison of the coefficients in front of $z_1^2E_{1+1}$ on
both sides of (\ref{maineq}) gives: \be G_{ij}-k^1_ik^1_j =
\Big(k^{12}_ik^{14}_j+k^{14}_ik^{12}_j\Big) z_2z_4(-{\bf p}_1{\bf
p}_4) = -\eta^2\Big(k^1_ik^1_j-k^2_ik^2_j\Big) \ee where $\eta^2 =
z_2z_4t_{24}$. Similarly, from the coefficient of $z_2^2E_{2+2}$ one obtains \be
G_{ij}-k^2_ik^2_j = \Big(k^{12}_ik^{23}_j+k^{23}_ik^{12}_j\Big)
z_1z_3({\bf p}_1{\bf p}_2) = \xi^2\Big(k^1_ik^1_j-k^2_ik^2_j\Big)
\ee with $\xi^2 = z_1z_3t_{13}$. Together these two equations imply
the consistency relation on the parameters $z_a$ \be
\xi^2+\eta^2=z_1z_3t_{13}+z_2z_4t_{24}=1\label{n=4constraint} \ee already familiar from
\cite{MMT}, and the explicit expression for $G_{ij}$, \be G_{ij} = \xi^2
k^1_ik^1_j + \eta^2 k^2_ik^2_j \label{Gthk} \ee All other relations,
following from (\ref{maineq}), are then automatically satisfied. For
example, the coefficient of $z_1z_2E_{1+2}$ on the right hand side
of (\ref{maineq}) receives contributions from $a+b+c+d=1+1+2+3$ and
$1+2+2+4$, and using the above relations one has \be \Big(k^{12}_ik^{13}_j+k^{13}_ik^{12}_j\Big)
z_1z_3\Big({\bf p}_1({\bf p}_1+{\bf p}_2)\Big) +
\Big(k^{12}_ik^{24}_j+k^{24}_ik^{12}_j\Big)
z_2z_4\Big({\bf p}_1({\bf p}_2+{\bf p}_3)\Big) =\nn\\
= \Big(2k^1_ik^1_j - k^1_ik^2_j-k^2_ik^1_j\Big)z_1z_3t_{13}
-\Big(2k^2_ik^2_j - k^1_ik^2_j-k^2_ik^1_j\Big)z_2z_4(-t_{24}) = \nn
\\ = 2(\xi^2 k^1_ik^1_j + \eta^2 k^2_ik^2_j)
-(k^1_ik^2_j+k^2_ik^1_j)(\xi^2+\eta^2)= \nn \\
=2G_{ij}-k^1_ik^2_j-k^2_ik^1_j \ee the last expression being equal to the coefficient of the
same term on the left hand side. Similarly for the coefficients of
$E_0=1$.

It remains to substitute $G_{ij}$ from (\ref{Gthk}) into
equations (\ref{keqS}) and (\ref{keqNG}).

In the $\sigma$-model case (\ref{keqS}) leads to \be \vec k_1^2 = \vec k_2^2 =
{\rm Tr}\ G = \xi^2 \vec k_1^2 + \eta^2 \vec k_2^2  \ee As soon as
the two vectors $\vec k_1$ and $\vec k_2$ have equal lengths,
the corresponding parallelogram has to be a rectangle. The
remaining essential modulus is the angle $\phi$ between the two
vectors, their common length being an inessential modulus (scaling
of the lagrangian). Another essential modulus is $\xi^2$ or
$\eta^2=1-\xi^2$. Rescalings of parameters $z_a$, which leave
$\xi^2$ and $\eta^2$ intact, are induced by constant shifts of the
coordinate vectors $\vec u$.

Analogously, in the NG case, one obtains from (\ref{keqNG}) \be \check
G^{ij}k_i^1k_j^1 = \check G^{ij}k_i^2k_j^2 = 2 \det G \label{NGsq}
\ee If we parametrize the two NG $\vec k$-vectors through $\vec
k_1=(\alpha,\beta)$ and $\vec k_2=(\gamma,\delta)$, then \be G_{ij}
= \left(\begin{array}{cc} \alpha^2\xi^2+\gamma^2\eta^2 &
\alpha\beta\xi^2+\gamma\delta\eta^2\\
\alpha\beta\xi^2+\gamma\delta\eta^2 &
\beta^2\xi^2+\delta^2\eta^2\end{array}\right), \ \ \ \ \
\check G^{ij} = \left(\begin{array}{cc}
\beta^2\xi^2+\delta^2\eta^2 &
-\alpha\beta\xi^2-\gamma\delta\eta^2\\
-\alpha\beta\xi^2-\gamma\delta\eta^2 &
\alpha^2\xi^2+\gamma^2\eta^2\end{array}\right),
\ee
$\det G = (\alpha\delta-\beta\gamma)^2\xi^2\eta^2$,
and (\ref{NGsq}) is equivalent to the system of equations
\be
\check G^{ij}k_i^1k_j^1 =
(\alpha^2\delta^2-2\alpha\beta\gamma\delta + \beta^2\gamma^2)\eta^2
= 2(\alpha\delta-\beta\gamma)^2\xi^2\eta^2, \nn\\
\check G^{ij}k_i^2k_j^2 = (\alpha^2\delta^2-2\alpha\beta\gamma\delta
+ \beta^2\gamma^2)\xi^2 = 2(\alpha\delta-\beta\gamma)^2\xi^2\eta^2
\ee
It follows that \be \xi_{NG}^2=\eta_{NG}^2=\frac{1}{2} \ee with no restriction on the
vectors $\vec k_1$ and $\vec k_2$. The corresponding parallelogram in this case
can be arbitrary
and the two essential moduli are the angle $\phi$ between $\vec k_1$ and $\vec
k_2$ and the ratio of their lengths, $|\vec k_1|/|\vec k_2|
=\sqrt{(\alpha^2+\beta^2)/(\gamma^2+\delta^2)}$. It is clear
that {\it arbitrary} vectors $\vec k_1$ and $\vec k_2$ give rise
to NG solutions, because they can be made arbitrary by linear transformations of $u^i$, which
are part of the $2d$ reparametrization invariance of the NG equations i.e. part
of the symmetry group of ${\cal M}_4^{NG}$.

\subsection{On the relation between NG and $\sigma$-model solutions}

In the previous section we determined the moduli spaces of both the $\sigma$-model
and the NG
equations for $n=4$ in the framework of the ansatz (\ref{exposol}).
Actually, the $\sigma$-model case corresponds to the trace of
equation (\ref{maineq}) with respect to the indices $i,j$ \cite{MMT}, while
in the  NG case one should contract with $\check G^{ij}$ instead of $\delta^{ij}$.

Our result is that the moduli spaces, while both
two-dimensional, are essentially different. Moreover, there is
no one-to-one correspondence between the solutions. Does this contradict the
widespread belief that the NG and the $\sigma$-model are equivalent? It does not,
as we shall argue next, because of the Virasoro constraints. Notice that the
$\sigma$-model dealt with in \cite{AM} and \cite{MMT} does not take into account the Virasoro
constraints, which are crucial in the proof of the above equivalence.
So, the two models are actually different and, not surprisingly, lead to different answers.

More specifically, recall that the idea
behind the equivalence of the NG and $\sigma$-model formalisms is based
on the consideration of the more general Polyakov action \cite{Pol}:
\be\label{Pol} \int {\cal L}_P\ d^2u = \int {\cal G}_{ab}g^{ij}\partial_i
X^a\partial_j X^b\sqrt{g}\ d^2u \ee where ${\cal G}_{ab}$ is the
target space metric, made from dynamical fields like in
(\ref{Gmet}), $X^a\equiv (r,{\bf v})$ and $g_{ij}$ is the auxiliary field
of $2d$-metric. The equations of motion for the dynamical fields then read
\be\label{eqmX}
\partial_i\Big(g^{ij}{\cal G}_{ab}\sqrt{g}\ \partial_j X^a\Big) =
{\partial {\cal L}_P\over\partial X^b} \ee while variation with respect to the $2d$-metric
gives \be g_{ij}=2{{\cal G}_{ab}\partial_i X^a\partial_j X^b\over {\cal
G}_{cd}g^{kl}\partial_k X^c\partial_l X^d}=2{G_{ij}\over
g^{kl}G_{kl}}\label{2dmetrics}\ee Inserting (\ref{2dmetrics}) into
(\ref{eqmX}), one reproduces the NG equations, while, taking advantage of the local symmetries
of the Polyakov action to choose
$g_{ij} = \delta_{ij}$, one obtains the $\sigma$-model equations. This
choice is a gauge-fixing and can always be achieved by a proper transformation of the world sheet
variables $u^i$ \footnote{It is well known that the freedom of arbitrary reparametrizations of the world
sheet is enough to render an arbitrary metric conformally flat; however, the
conformal factor is inessential due to the Weyl invariance of the
action, $g_{ij}\to\rho g_{ij}$.}.
Based on this, one may argue that any solution of the NG
equations can be converted into a solution of the $\sigma$-model: once a
$G^{NG}_{ij}$ is found, it can always be diagonalized
by a coordinate transformation. In our context, with $G_{ij}$ constant,
this transformation $\vec u^{NG} \rightarrow \vec u^\sigma$
is linear, and is given simply by
\be
\vec k_a^{NG}\vec u^{NG} = \vec k_a^\sigma \vec u^\sigma,
\ee
or equivalently, using the explicit form of $k_a^{NG}$ and $k_a^\sigma$,
\be
u_1^\sigma = \alpha u_1^{NG} + \beta u_2^{NG}, \nn \\
u_2^\sigma = \gamma u_1^{NG} + \delta u_2^{NG}
\ee
It is always possible to find such a transformation with non-unit Jacobian,
in order  to convert the two NG $\vec k$-vectors with different lengths into
two $\sigma$-model $\vec k$-vectors with equal lengths.
Clearly, the above $NG \rightarrow \sigma$ mapping has a non-trivial kernel. It has enough
parameters to map different NG solutions into the same $\sigma$-model
solution; it is not an isomorphism of the two moduli spaces.

The converse, however, is not true: one cannot convert an arbitrary $\sigma$-model solution
into a NG one. For this, one would have in addition to satisfy the gauge condition
$G_{ij}\sim\delta_{ij}$. For instance, linear transformations of
coordinates $\vec u$ cannot change parameters $\xi^2$ and $\eta^2$.
The parameters $z_a$ of a particular solution (\ref{exposol})
 are rescaled by {\it shifts} of $\vec u$, $\vec u
\rightarrow \vec u + \vec a$, but $z_1$ and $z_3=z_{-1}$ or $z_2$
and $z_4=z_{-2}$ are rescaled in opposite directions (since $\vec
k_{-a} = -\vec k_a$), so that $\xi^2$ and $\eta^2=1-\xi^2$ remain
intact. This implies that it is not possible to use the gauge freedom of the Polyakov
equations to convert $\sigma$-model
solutions with generic $\xi^2\neq 1/2$ into NG solutions, which all
have $\xi^2=\eta^2=1/2$. Generic coordinate-$\vec u$ reparametrizations (linear or otherwise)
change the two tensors $g_{ij}$ and
$G_{ij}$ simultaneously, and the desired transformation $\Big(
g^\sigma_{ij}, G^\sigma_{ij}\Big) \stackrel{?}{\longrightarrow}
\Big( g^{NG}_{ij}, G^{NG}_{ij}\Big)$ is generically in
contradiction with the other two properties, namely \be g^\sigma_{ij} =
\delta_{ij} \ee and (\ref{2dmetrics}) \be g^{NG}_{ij} \sim G^{NG}_{ij}\ee These
relations are all compatible if and only if $G^\sigma_{ij}
\sim\delta_{ij}$, which is {\it not} true for a generic $\sigma$-model
solution, but only for those with $\xi_\sigma^2=\eta_\sigma^2=1/2$. This, as stated in
the begining of this section, is a concrete manifestation of the well known fact \cite{vico} that
Polyakov's $\sigma$-model, which is classically equivalent to the NG theory,
reproduces the ordinary $\sigma$-model, but together with the Virasoro
constraints.

A consequence of the above discussion is that the regularized $\sigma$-model and NG
actions of even a common solution do not coincide. Naively, since substitution
of the $2d$-metric (\ref{2dmetrics}) into the Polyakov action (\ref{Pol})
reproduces the NG action, one would expect that the $\sigma$-model
and NG actions coincide, provided the
Virasoro constraint is satisfied. This is true, but ambiguous, since both actions
are infinite. The regularization proposed in
\cite{AM} does not change only the target space metric ${\cal G}_{ab}$ in
both actions, which would leave them equal. Instead, it spoils the Virasoro constraint
and should lead a priori to different actions! Indeed, it was explicitly checked \cite{Popo} that,
even in the $n=4$ case, the two actions are different. However, they differ by an inessential additive
constant. It would be instructive to examine their difference for
higher $n$.


\section{Guess of the action integral for $n=5$}

As explained in \cite{MMT}, it is not straightforward to generalize to $n>4$ our solutions with exponential
behavior at infinity.
So, it is not obvious how to extend our approach to these cases and have so far been unable
to find relevant solutions.
Nevertheless, one can still try to {\it guess} the form of the regularized action integral
${\cal A}_n(z_1,...,z_n;\epsilon)$ for $n\geq 5$, whose minimum will lead to the BDDK formula for the
one-loop amplitude $F_n^{(1)}$ of n-gluon scattering.
For that, let us {\it assume} that we have an n-parameter set of solutions of the $\sigma$-model
with the appropriate asymptotics, parametrized by $z_a$, with $a=1,2,...,n$. In addition, we must
assume a regularization scheme \cite{AM} with parameter $\epsilon$, as well as a constraint
analogous to (\ref{n=4constraint}).

We split the action integral into the infinite ${\cal A}^{(n)}_\epsilon$ and finite $\tilde{\cal A}_n$
pieces and guided by the pictorial representation of the BDDK formula \cite{MMT} and by our $n=4$
results, we write (up to an additive inessential constant $1/\epsilon^2$ term)
\be {\cal A}_n(z_1,z_2,...,z_n)={\cal A}^{(n)}_\epsilon + \tilde{\cal A}_n
= \frac{1}{\epsilon}\sum_{a=1}^n\log z_a + \sum_{a=1}^n \log z_a \log z_{a+1} \label{Aguess}\ee
with $a+n\equiv a$ for all values of $a$. We
neglected any additional angular variables like $\phi$ of the $n=4$ case, assuming that such parameters enter in an
especially simple way, like it happened in the case of $n=4$ in
\cite{MMT}, where $|\sin\phi|^{-1}$ was a common factor in front of
the entire ${\cal A}_4$. Notice that for $n=4$ (\ref{Aguess}) reproduces the expression derived
in \cite{MMT}.

To guess a reasonable generalization of the constraint is more difficult.
For general $n$ one has to worry about the presence of terms with higher powers of $t_{ab}$
in the expression for the constraint. For instance, the first non-trivial
such term would be $\sum_{a<b<c<d} z_az_bz_cz_dt_{ab}t_{cd}$. However, for $n=4, 5$
such a term, as well as all analogous with higher powers of $t$ vanish identically.
In what follows, we shall consider the constraint
\be \sum_{a<b}^n z_az_bt_{ab}=1 \label{constraint}\ee but, one should remember, that this particular
form may be oversimplified and irrelevant for $n\geq 6$.

Our goal is to minimize (\ref{Aguess}) under the constraint (\ref{constraint}). Let us
start with the simpler problem of minimizing ${\cal A}^{(n)}_\epsilon$ under the
above constraint, which is introduced with a lagrange multiplier $\lambda$.
The position $z_a^{(0)}$ of the minimum satisfies
\be\label{min} \frac{1}{z_a^{(0)}} =
\lambda \sum_{b=1}^n t_{ab} z_b^{(0)} \ee

For $n=4$, the solution is, up to the invariance $z_1^{(0)}\to\zeta z_1^{(0)}$,
$z_3^{(0)}\to {1\over\zeta}z_3^{(0)}$ and $z_2^{(0)}\to\zeta' z_2^{(0)}$, $z_4^{(0)}\to
{1\over\zeta'}z_4^{(0)}$, given by \cite{MMT} \be\label{sol4} z_a^{(0)} =
\frac{1}{\sqrt{2t_{a,a+2}}} \ee

Similarly, for $n=5$ we obtain instead \be z_a^{(0)}z_b^{(0)}
={1\over 5t_{ab}},\ \ \ \ \ \lambda={5\over 2}\ee Multiplying all
these pairs together gives \be
z_1^{(0)}z_2^{(0)}z_3^{(0)}z_4^{(0)}z_5^{(0)}={1\over\sqrt{5^5t_{13}t_{14}t_{23}t_{24}t_{35}}} \ee
Now, dividing this expression twice by appropriately chosen products $z_a^{(0)}z_b^{(0)}$,
one obtains \be\label{sol5} z_a^{(0)}= \sqrt{\frac{t_{a+1,a+3}t_{a+2,a+4}}
{5t_{a,a+2}t_{a,a+3}t_{a+1,a+4}}}\label{minz} \ee If one denotes $\tau_{ab}=\log t_{ab}$,
(\ref{minz}) has a pictorial representation shown in Figure \ref{picminz}.
Incidentally, note that in contrast to the $n=4$ case, there is no rescaling
freedom in solutions of equation (\ref{min}) for $n=5$.

Going back to the minimization of ${\cal A}$, observe that the presence of the finite correction
$\tilde{\cal A}_n$ in ${\cal A}$ will shift the position of the minimum to
$z_a=z_a^{(0)}+\epsilon z_a^{(1)}$. The ${\cal O}(\epsilon)$ $z_a^{(1)}$-shift of $z_a$ could a priori give
a finite correction to ${\cal A}$. However, as we will
argue, this ${\cal O}(\epsilon^0)$ contribution actually vanishes.
Indeed, the finite correction of ${\cal A}$ due to
$z_a^{(1)}$ is \be \left.\sum_a^n{\partial{\cal A}_\epsilon\over
\partial z_a}\right|_{z_a=z_a^{(0)}}z_a^{(1)}=\sum_a^n{z_a^{(1)}\over
z_a^{(0)}}=\lambda\sum_{a,b}^n t_{ab}z_a^{(0)}z_b^{(1)}=0 \ee the
last two equalities being direct corollaries of (\ref{min}).

Thus, in order to reproduce the BDDK\footnote{In this particular
case of $n=5$ it should rather be named BDK formula \cite{BDK}, this example was
actually used as a basis for further calculations at $n\geq 6$ in
\cite{BDDK} and structures revealed at $n=5$ are inherited by
generic BDDK expressions. The BDK formula has a number of equivalent
representations, of which we use just one in this paper.} result, one has to insert
the solutions (\ref{minz})
for $z_a^{(0)}$ into the action (\ref{Aguess}).
The result is the BDK formula for $n=5$ (in this formula one should
put $\mu^2=1/5$),  \be BDK\ =\ BDDK_5\ = \left.-{1\over\epsilon^2}
\prod_a\left({\mu^2\over
t_{a,a+2}}\right)^\epsilon+\sum_a\log{t_{a,a+2}\over
t_{a+1,a+3}}\log{t_{a+2,a+4}\over t_{a-2,a}}=-2{\cal A}_5
\right|_{z_a=z_a^{(0)}}+O(\epsilon) \ee

The finite part of this
expression is equal to (see Figure \ref{picBDS5}) \be
\sum_{a=1}^n \Big(\tau_{a,a+2} - \tau_{a,a-2}\Big)
\Big(\tau_{a-1,a+2} - \tau_{a+1,a-2}\Big) =
(\tau_{14}-\tau_{13})(\tau_{24}-\tau_{35}) + {\rm cyclic\
permutations}, \label{BDS5} \ee This generalizes the older result
for $n=4$ \be \hbox{finite part of }BDDK_4\ =
\Big(\tau_{13}-\tau_{24}\Big)^2 = \left(\log\frac{s}{t}\right)^2 \ee

\begin{figure}
\unitlength 1mm 
\linethickness{0.4pt}
\ifx\plotpoint\undefined\newsavebox{\plotpoint}\fi 
\begin{picture}(140.027,72.273)(0,0)
\multiput(48.794,31.878)(-.0413822526,-.0337030717){293}{\line(-1,0){.0413822526}}
\multiput(48.153,70.394)(-.0413822526,-.0337030717){293}{\line(-1,0){.0413822526}}
\multiput(93.241,31.878)(-.0413822526,-.0337030717){293}{\line(-1,0){.0413822526}}
\multiput(92.6,70.394)(-.0413822526,-.0337030717){293}{\line(-1,0){.0413822526}}
\multiput(36.669,22.003)(.033717105,-.086348684){152}{\line(0,-1){.086348684}}
\multiput(36.028,60.519)(.033717105,-.086348684){152}{\line(0,-1){.086348684}}
\multiput(81.116,22.003)(.033717105,-.086348684){152}{\line(0,-1){.086348684}}
\multiput(80.475,60.519)(.033717105,-.086348684){152}{\line(0,-1){.086348684}}
\put(41.794,8.878){\line(1,0){15}}
\put(41.153,47.394){\line(1,0){15}}
\put(86.241,8.878){\line(1,0){15}}
\put(85.6,47.394){\line(1,0){15}}
\multiput(56.794,8.878)(.03362069,.095689655){145}{\line(0,1){.095689655}}
\multiput(56.153,47.394)(.03362069,.095689655){145}{\line(0,1){.095689655}}
\multiput(101.241,8.878)(.03362069,.095689655){145}{\line(0,1){.095689655}}
\multiput(100.6,47.394)(.03362069,.095689655){145}{\line(0,1){.095689655}}
\multiput(61.669,22.753)(-.0424354244,.0336715867){271}{\line(-1,0){.0424354244}}
\multiput(61.028,61.269)(-.0424354244,.0336715867){271}{\line(-1,0){.0424354244}}
\multiput(106.116,22.753)(-.0424354244,.0336715867){271}{\line(-1,0){.0424354244}}
\multiput(105.475,61.269)(-.0424354244,.0336715867){271}{\line(-1,0){.0424354244}}
\put(49.381,32.134){\circle{1.768}}
\put(48.74,70.649){\circle{1.768}}
\put(93.828,32.134){\circle{1.768}}
\put(93.187,70.649){\circle{1.768}}
\put(52.156,72.124){\makebox(0,0)[cc]{1}}
\put(96.751,72.273){\makebox(0,0)[cc]{1}}
\put(97.197,33.326){\makebox(0,0)[cc]{1}}
\put(52.602,33.772){\makebox(0,0)[cc]{1}}
\put(63.008,61.867){\makebox(0,0)[cc]{2}}
\put(107.603,62.016){\makebox(0,0)[cc]{2}}
\put(108.049,23.069){\makebox(0,0)[cc]{2}}
\put(63.454,23.515){\makebox(0,0)[cc]{2}}
\put(57.656,45.664){\makebox(0,0)[cc]{3}}
\put(102.251,45.813){\makebox(0,0)[cc]{3}}
\put(102.697,6.866){\makebox(0,0)[cc]{3}}
\put(58.102,7.312){\makebox(0,0)[cc]{3}}
\put(39.818,44.474){\makebox(0,0)[cc]{4}}
\put(84.413,44.623){\makebox(0,0)[cc]{4}}
\put(84.859,5.676){\makebox(0,0)[cc]{4}}
\put(40.264,6.122){\makebox(0,0)[cc]{4}}
\put(33.426,59.488){\makebox(0,0)[cc]{5}}
\put(78.021,59.637){\makebox(0,0)[cc]{5}}
\put(78.467,20.69){\makebox(0,0)[cc]{5}}
\put(33.872,21.136){\makebox(0,0)[cc]{5}}
\put(10.831,60.231){\makebox(0,0)[cc]{$\log z_1=$}}
\put(8.027,20.811){\makebox(0,0)[cc]{$\left(BDDK_5\right)_{f.p.}=$}}
\put(26.407,25.675){\line(-1,0){6.096}}
\multiput(20.31,25.675)(.044654867,-.033486726){113}{\line(1,0){.044654867}}
\multiput(25.356,21.891)(-.043725664,-.033486726){113}{\line(-1,0){.043725664}}
\put(20.415,18.107){\line(1,0){6.307}}
\put(26.722,18.107){\line(0,1){.736}}
\put(26.407,24.729){\line(0,1){1.051}}
\put(23.884,15.689){\circle{1.784}}
\put(71.352,21.257){\circle*{1.994}}
\put(68.482,60.516){\line(1,0){4.757}}
\thicklines
\multiput(93.603,31.113)(-.033610329,-.103741784){213}{\line(0,-1){.103741784}}
\multiput(92.839,70.086)(-.033610329,-.103741784){213}{\line(0,-1){.103741784}}
\multiput(93.515,31.024)(-.033711628,-.101953488){215}{\line(0,-1){.101953488}}
\multiput(92.751,69.998)(-.033711628,-.101953488){215}{\line(0,-1){.101953488}}
\multiput(93.338,31.201)(-.033610329,-.102497653){213}{\line(0,-1){.102497653}}
\multiput(92.574,70.175)(-.033610329,-.102497653){213}{\line(0,-1){.102497653}}
\multiput(93.957,31.201)(.033711628,-.104009302){215}{\line(0,-1){.104009302}}
\multiput(93.193,70.175)(.033711628,-.104009302){215}{\line(0,-1){.104009302}}
\multiput(94.222,31.201)(.033671429,-.10522381){210}{\line(0,-1){.10522381}}
\multiput(93.458,70.175)(.033671429,-.10522381){210}{\line(0,-1){.10522381}}
\multiput(94.31,31.289)(.033671429,-.1048){210}{\line(0,-1){.1048}}
\multiput(93.546,70.263)(.033671429,-.104804762){210}{\line(0,-1){.104804762}}
\multiput(41.984,8.839)(.0485717822,.0336930693){404}{\line(1,0){.0485717822}}
\multiput(41.072,47.516)(.0485717822,.0336905941){404}{\line(1,0){.0485717822}}
\multiput(61.607,22.716)(-.0485717822,-.0336930693){404}{\line(-1,0){.0485717822}}
\multiput(60.695,61.393)(-.0485717822,-.0336930693){404}{\line(-1,0){.0485717822}}
\multiput(61.518,22.804)(-.048408313,-.0337114914){409}{\line(-1,0){.048408313}}
\multiput(60.606,61.481)(-.048408313,-.0337139364){409}{\line(-1,0){.048408313}}
\multiput(36.77,21.744)(.0525657895,-.0337289474){380}{\line(1,0){.0525657895}}
\multiput(35.857,60.42)(.0525684211,-.0337263158){380}{\line(1,0){.0525684211}}
\multiput(36.77,21.92)(.052845953,-.033691906){383}{\line(1,0){.052845953}}
\multiput(35.857,60.597)(.052848564,-.033694517){383}{\line(1,0){.052848564}}
\multiput(56.834,9.016)(-.0528465608,.0336719577){378}{\line(-1,0){.0528465608}}
\multiput(55.922,47.692)(-.0528465608,.0336719577){378}{\line(-1,0){.0528465608}}
\multiput(80.548,60.529)(1.3791667,.033){18}{\line(1,0){1.3791667}}
\multiput(80.994,60.826)(1.7307143,.0318571){14}{\line(1,0){1.7307143}}
\multiput(105.224,60.826)(-2.725222,-.033){9}{\line(-1,0){2.725222}}
\put(139.831,60.231){\makebox(0,0)[cc]
{$= \tau_{24}+\tau_{35}-\tau_{13}-\tau_{14}-\tau_{25}$}}
\put(140.027,20.811){\makebox(0,0)[cc]
{$=(\tau_{24}-\tau_{35})\cdot(\tau_{14}-\tau_{13}) + {\rm perms}$}}
\thinlines
\multiput(7.284,61.541)(.1486,.0298){5}{\line(1,0){.1486}}
\end{picture}
\caption{{\footnotesize Pictorial representations of
eqs.(\ref{minz}) and (\ref{BDS5}). The polygon is nothing but $\Pi$
from Figure \ref{polyg}, whose edges are associated with external
momenta. $\tau$-parameters $\tau_{ab} = \log t_{ab}$ are associated
with diagonals and $z_a$ with the vertices of the polygon. The
marked diagonals correspond to differences of $\tau$-parameters in
the equations.}} \label{picminz} \label{pictildA} \label{picBDS5}
\end{figure}

It is easy to see that the expressions for $n=5$ are natural
generalizations of those for $n=4$. The main new ingredient for $n>5$ is
that $t^{[r]}$ with $r>2$ (see \cite{MMT} for notations) and higher
powers of $t$ can enter the constraint. At the same time,
dilogarithmic contributions should appear in the action integral. It
can happen that they occur after additional integration over some
new moduli. We do not go into details of these subtler considerations
in the present text.

\section*{Acknowledgements}

We are indebted to T.Mironova for help with the pictures.
Work is supported in part by the INTERREG-IIIA Greece-Cyprus
program, as well as by the European Union contract MRTN-CT-512194.
A.Mironov and A.Morozov acknowledge the kind hospitality and support
of the Institute of Plasma Physics and the Department of Physics of
the University of Crete during the work on this paper.
A.Morozov also acknowledges the hospitality of ESI, Vienna,
at the last stage.
The work of A.M.'s is partly supported by Russian Federal Nuclear
Energy Agency, by the joint grant 06-01-92059-CE,  by NWO project
047.011.2004.026, by INTAS grant 05-1000008-7865, by
ANR-05-BLAN-0029-01 project and by the Russian President's Grant of
Support for the Scientific Schools NSh-8004.2006.2, by RFBR grants
07-02-00878 (A.Mir.) and 07-02-00645 (A.Mor.).


\begin{thebibliography}{12}

\bibitem{AM} L.Alday and J.Maldacena, {\it Gluon Scattering Amplitudes at Strong
Coupling}, arXiv:0705.0303

\bibitem{BDS} Z.Bern, L.Dixon and V.Smirnov,
{\it Iteration of Planar Amplitudes in Maximally Supersymmetric
Yang-Mills Theory at Three Loops and Beyond}, Phys.Rev. {\bf D72}
(2005) 085001, hep-th/0505205

\bibitem{AMoth} S.Abel, S.Forste and V.Khose, {\it Scattering Amplitudes in Strongly
Coupled $N=4$ SYM from Semiclassical Strings in AdS},
arXiv:0705.2113

\bibitem{AMoth2} E.Buchbinder, {\it Infrared Limit of Gluon Amplitudes at Strong
Coupling}, arXiv:0706.2015

\bibitem{AMoth3} J.Drummond, G.Korchemsky and E.Sokatchev,
{\it Conformal properties of four-gluon planar amplitudes and Wilson
loops}, arXiv:0707.0243

\bibitem{AMoth4}
A.Brandhuber, P.Heslop and G.Travaglini, {\it MHV Aplitudes in $N=4$
Super Yang-Mills and Wilson Loops}, arXiv:0707.1153

\bibitem{AMoth5} F.Cachazo,
M.Spradlin and A.Volovich, {\it  Four-Loop Collinear Anomalous
Dimension in N = 4 Yang-MillsTheory}, arXiv:0707.1903

\bibitem{AMoth6} M.Kruczenski,
R.Roiban, A.Tirziu and A.Tseytlin, {\it Strong-Coupling Expansion of
Cusp Anomaly and Gluon Amplitudes from Quantum Open Strings in
$AdS_5\times S^5$}, arXiv:0707.4254

\bibitem{AMoth7} Z.Komargodsky and S.Razamat,
{\it  Planar Quark Scattering at Strong Coupling and Universality},
arXiv:0707.4367

\bibitem{AMoth8} L.Alday and J.Maldacena, {\it Comments on Operators
with Large Spin}, arXiv:0708.0672; {\it  Comments on gluon
scattering amplitudes via AdS/CFT}, arXiv:0710.1060

\bibitem{AMoth9} A.Jevicki,
C.Kalousios, M.Spradlin and A.Volovich, {\it Dressing the Giant
Gluon}, arXiv:0708.0818

\bibitem{MMT} A.Mironov, A.Morozov and T.N.Tomaras, {\it On n-point Amplitudes
in N=4 SYM}, arXiv:0708.1625

\bibitem{AMoth10} H.Kawai and T.Suyama, {\it Some Implications
of Perturbative Approach to AdS/CFT Correspondence}, arXiv:0708.2463

\bibitem{AMoth11}
S.G.Naculich and H.J.Schnitzer, {\it  Regge behavior of gluon
scattering amplitudes in N=4 SYM theory}, arXiv:0708.3069

\bibitem{AMoth12} R.Roiban
and A.A.Tseytlin, {\it  Strong-coupling expansion of cusp anomaly
from quantum superstring}, arXiv:0709.0681

\bibitem{AMoth13} D.Nguyen, M.Spradlin and A.Volovich, {\it  New Dual Conformally
Invariant Off-Shell Integrals}, arXiv:0709.4665

\bibitem{AMoth14} J.McGreevy and
A.Sever, {\it  Quark scattering amplitudes at strong coupling},
arXiv:0710.0393

\bibitem{AMoth15} S.Ryang, {\it  Conformal SO(2,4) Transformations of
the One-Cusp Wilson Loop Surface}, arXiv:0710.1673

\bibitem{AMoth16} D.Astefanesei,
S.Dobashi, K.Ito and H.S.Nastase, {\it  Comments on gluon 6-point
scattering amplitudes in N=4 SYM at strong coupling},
arXiv:0710.1684

\bibitem{Popo} A.Popolitov, {\it On coincidence of
Alday-Maldacena-regularized $\sigma$-model and Nambu-Goto areas of
minimal surfaces}, arXiv:0710.2073

\bibitem{BDK} Z.Bern, L.Dixon and D.A.Kosower, {\it One-Loop Corrections
to Five-Gluon Amplitudes}, Phys.Rev.Lett. {\bf 70} (1993) 2677-2680,
hep-ph/9302280; {\it Dimensionally Regulated Pentagon Integrals},
Nucl.Phys. {\bf B412} (1994) 751-816, hep-ph/9306240

\bibitem{BDDK} Z.Bern, L.Dixon, D.C.Dunbar and D.A.Kosower, {\it One-Loop n-Point
Gauge Theory Amplitudes, Unitarity and Collinear Limits}, Nucl.Phys.
{\bf B425} (1994) 217-260, hep-ph/9403226; {\it One-Loop Gauge
Theory Amplitudes with an Arbitrary Number of External Legs},
hep-ph/9405248

\bibitem{Pol} A.Polyakov, {\it Quantum Geometry of Bosonic Strings},
Phys.Lett. {\bf B103} (1981) 207-210

\bibitem{vico} A.Polyakov, {\sl  Gauge Fields And Strings}, 1987\\
M.B.Green, J.H.Schwarz and E.Witten, {\sl Superstring Theory},
v.1-2, Cambridge University Press, 1987

\end{thebibliography}
\end{document}